\documentclass[12pt]{iopart}
\usepackage{iopams}
\usepackage{setstack}

\newcommand{\half}{\mbox{$\textstyle \frac{1}{2}$}}
\newcommand{\cP}{{\cal P}}
\newcommand{\cT}{{\cal T}}

\newcommand{\cQ}{{\cal Q}}

\begin{document}


\title[Interactions of Hermitian and non-Hermitian Hamiltonians]
{Interactions of Hermitian and non-Hermitian Hamiltonians}

\author[Bender and Jones]{Carl~M~Bender${}^\ast$\footnote{Permanent address:
Department of Physics, Washington University, St. Louis MO 63130, USA;
\\{\footnotesize{\tt email: cmb@wustl.edu}}} and Hugh
F.~Jones${}^\dag$\footnote{{\footnotesize{\tt email:
h.f.jones@imperial.ac.uk}}}}

\address{${}^\ast$Center for Nonlinear Studies, Los Alamos National Laboratory,
Los Alamos, NM 87545, USA}

\address{${}^\dag$Blackett Laboratory, Imperial College, London SW7 2BZ, UK}

\date{today}

\begin{abstract}
The coupling of non-Hermitian $\cP\cT$-symmetric Hamiltonians to standard
Hermitian Hamiltonians, each of which individually has a real energy spectrum,
is explored by means of a number of soluble models. It is found that in all
cases the energy remains real for small values of the coupling constant, but
becomes complex if the coupling becomes stronger than some critical value. For a
quadratic non-Hermitian $\cP\cT$-symmetric Hamiltonian coupled to an arbitrary
real Hermitian $\cP\cT$-symmetric Hamiltonian, the reality of the ground-state
energy for small enough coupling constant is established up to second order in
perturbation theory.
\end{abstract}

\submitto{\JPA}

\section{Introduction}
\label{Intro}
Since the work by Bender and Boettcher \cite{BB} on non-Hermitian but $\cP
\cT$-symmetric Hamiltonians, subsequent research has gone through various
stages. First came an exploration of various non-Hermitian generalizations of
well-known soluble models to determine if their spectra is real. However,
reality of the spectrum does not by itself guarantee a viable quantum theory.
One also needs a probabilistic interpretation, and since the most obvious choice
of metric for a $\cP\cT$-symmetric model is not positive definite, a Hilbert
space endowed with this metric does not represent a physical framework for
quantum mechanics. Instead, one must find an alternative, positive-definite
metric \cite{BBJ,Gey,AM-metric}, which is dynamically determined by the
particular Hamiltonian in question. It was further shown \cite{AM-h} that this
metric $\eta\equiv e^{-Q}$ provides a similarity transformation from the
non-Hermitian $H$ to an equivalent Hermitian $\tilde{H}$. This equivalent
Hermitian Hamiltonian was subsequently constructed, often in perturbation theory
only, in a variety of models \cite{HFJ-ix3,AM-ix3,-x4}.

These investigations were all concerned with isolated non-Hermitian systems, but
more recently much attention has been given to situations where a non-Hermitian
system interacts with the world of Hermitian quantum mechanics. For example,
Ref.~\cite{QB} examined a non-Hermitian analogue of the Stern-Gerlach experiment
in which the role of the intermediate inhomogeneous magnetic field flipping the
spin is replaced by an apparatus described by a non-Hermitian Hamiltonian.
This type of set-up has been further elaborated by Assis and Fring \cite{Fri}
and G\"unther {\it et al}.~\cite{Uwe}, and it continues to be the focus of
lively discussion \cite{Martin,AM-QB,GS,Rot,AM1}. Recently, Ref.~\cite{HFJ}
explored the problem of scattering from localized non-Hermitian potentials.

It is in this spirit that we investigate the nature of the energy spectrum when
Hermitian and non-Hermitian systems, individually having real spectra, are
coupled together. In Sec.~\ref{s2} we first look at a simple matrix model and
then in Sec.~\ref{s3} we explore couplings of various non-Hermitian quadratic
Hamiltonians to a simple harmonic oscillator. Section \ref{s4} examines the
reality of the spectrum in perturbation theory for a complex quadratic $\cP
\cT$-symmetric Hamiltonian coupled to a generic real $\cP\cT$-symmetric {\it
and} Hermitian Hamiltonian. We summarize our results in Sec.~\ref{s5}.

\section{Simple Matrix Model}
\label{s2}

In this section we consider coupling a Hermitian matrix Hamiltonian
\begin{equation}
H_1=a\mathbf{1}+b\sigma_1=\left(\begin{array}{cc}a&b\\b&a\end{array}\right)
\quad(a,\,b~{\rm real})
\label{e1}
\end{equation}
to the non-Hermitian $\cP\cT$-symmetric matrix Hamiltonian introduced in
Ref.~\cite{BBJ}
\begin{equation}
H_2=r(\mathbf{1}\cos\theta +i\sin\theta\sigma_3)+s\sigma_1=\left(\begin{array}{cc}re^{i
\theta}& s\\s & re^{-i\theta}\end{array}\right)\quad(r,\,s,\,\theta~{\rm real}),
\label{e2}
\end{equation}
where $\mathbf{1}$ is the identity matrix and $\sigma_k$ are the Pauli matrices.
We choose the parameters $r$, $s$, and $\theta$ so that the inequality $s^2>r^2
\sin^2\theta$ is satisfied; this inequality guarantees that the eigenvalues of
$H_2$ are real \cite{BBJ}. The parity operator is taken as
\begin{equation}
\cP=\sigma_1=\left(\begin{array}{cc} 0\ 1\\ 1\ 0 \end{array}\right),
\label{e3}
\end{equation}
while $\cT$ implements complex conjugation. Note that both $H_1$ and $H_2$ are
symmetric under $\cP\cT$, and each separately has real eigenvalues.

To couple these two systems together we take the direct sum, but introduce
nonzero elements in the off-diagonal sector:
\begin{equation}
H=\left(\begin{array}{cc|cc}
a & b & \epsilon A & \epsilon B\\
b & a & \epsilon B^* & \epsilon A^* \\
\hline
\epsilon C & \epsilon D & re^{i\theta} & s\\
\epsilon D^* & \epsilon C^* & s & re^{-i\theta}\\
\end{array}
\label{e4}
\right)
\end{equation}
These are chosen in such a way that $H$ remains invariant under combined parity
reflection and time reversal, where the parity operator is given by
\begin{equation}
\cP={\mathbf 1}\otimes\sigma_1=\left(\begin{array}{cc|cc} 0 & 1 & 0 & 0\\
1 & 0 & 0 & 0\\ \hline 0 & 0 & 0 & 1\\ 0 & 0 & 1 & 0\\ \end{array}\right)
\label{e5}
\end{equation}
and time reversal is complex conjugation.

The question is whether the eigenvalues of this combined system remain real, and
if there is any constraint on the strength of the coupling parameter $\epsilon$.
As a specific example we choose $a=b=1$ in $H_1$, $r=s=1$ and $\theta=\pi/6$ in
$H_2$, and $A=C=1$, $B=D=0$ in the coupling matrices. Numerically we find that
in this case the eigenvalues remain real until $\epsilon$ exceeds a critical
value of approximately 0.7045. For other choices of the parameter the picture
is similar: In all cases the eigenvalues remain real for $\epsilon$ sufficiently
small. In some cases they appear to remain real for all values of $\epsilon$; in
others, as in the example above, they first become complex but then become real
again at a second critical value of $\epsilon$.

\section{Coupled Quadratic Hamiltonians}
\label{s3}

To determine if the energy levels of a coupled system of Hamiltonians are real,
our strategy here will be to find the $Q$ operator by using the condition that
\begin{equation}
H^\dag=e^{-\cQ}He^{\cQ},
\label{e6}
\end{equation}
and then to construct the equivalent Hermitian Hamiltonian $\tilde{H}$ by using
the similarity transformation
\begin{equation}
\tilde{H}=e^{-\cQ/2}He^{\cQ/2}.
\label{e7}
\end{equation}
In some cases the resulting $\tilde{H}$ will need to be diagonalized by a
further unitary transformation in order to identify the spectrum.

\subsection{Simple Harmonic Oscillator Coupled to a Shifted Simple Harmonic
Oscillator}

In this subsection we consider a quantum system described by the interaction of
a conventional and a $\cP\cT$-symmetric Hamiltonian:
\begin{equation}
H=(p^2+x^2)+(q^2+y^2+2iy)+2\epsilon xy.
\label{e8}
\end{equation}
Since the Hamiltonian (\ref{e8}) is quadratic, we expect the $\cQ$ operator to
be linear in the momentum variables,
\begin{equation}
{\cQ}=\alpha p+\beta q,
\label{e9}
\end{equation}
which will produce the coordinate shifts
\begin{eqnarray}
x\to x+i\alpha\quad{\rm and}\quad y\to y+i\beta.
\label{e10}
\end{eqnarray}

We determine $\alpha$ and $\beta$ by the condition (\ref{e6}), which gives
\begin{equation}
\alpha=\frac{\epsilon}{1-\epsilon^2}\quad\hbox{and}\quad\beta=-\frac{1}{1-
\epsilon^2}.
\label{e11}
\end{equation}
It is somewhat surprising that we are able to determine $\alpha$ and $\beta$
because the condition $H^\dag=e^{-\cQ}He^{\cQ}$ translates into a system of
three coupled linear equations with only two unknowns $\alpha$ and $\beta$. Yet,
there is a unique solution. However, note that the solution becomes {\it
singular} as $|\epsilon|$ reaches $1$.

Given $\cQ$ we construct $\tilde{H}$ according to Eq.~(\ref{e7}), which in this
case produces
\begin{equation}
\tilde{H}=e^{-\cQ/2}He^{\cQ/2}=p^2+x^2+q^2+y^2+2\epsilon xy+\frac{1}{1-
\epsilon^2}.
\label{e12}
\end{equation}

We have identified the equivalent Hermitian Hamiltonian $\tilde{H}$, but it
must still be diagonalized. To do so we change variables from $p$, $q$, $x$, and
$y$ to $P$, $Q$, $X$, and $Y$:
\begin{eqnarray}
p&=&aP+bQ,\nonumber\\ q&=&cP+dQ,\nonumber\\ x&=&eX+fY,\nonumber\\
y&=&gX+hY.
\label{e13}
\end{eqnarray}
We determine the unknown coefficients $a$ through $h$ by requiring that (i) the
canonical commutation relations
\begin{equation}
[p,q]=0,~~[x,y]=0,~~[y,p]=0,~~[x,q]=0,~~[x,p]=i,~~[y,q]=i
\label{e14}
\end{equation}
be maintained, and that (ii) $\tilde{H}$, when expressed in terms of $X$, $Y$,
$P$, and $Q$ contains no crossterms.

These two sets of conditions translate into six equations for the coefficients.
The solutions to these six equations are:
\begin{equation}
c=\eta a,\quad d=-\eta b,\quad e=\frac{1}{2a},\quad f=\frac{1}{2b},\quad g=
\frac{\eta}{2a},\quad h=-\frac{\eta}{2b},
\label{e15}
\end{equation}
where $\eta=\pm1$ and $a$ and $b$ are arbitrary. The resulting $\tilde{H}$ is
given by
\begin{equation}
\tilde{H}=2a^2P^2+\frac{1+\eta\epsilon}{2a^2}X^2+2b^2Q^2+\frac{1-\eta\epsilon}{2
b^2}Y^2+\frac{1}{1-\epsilon^2},
\label{e16}
\end{equation}
which is the sum of two uncoupled quantum-harmonic-oscillator Hamiltonians.

Since the energy levels of the general quantum harmonic oscillator Hamiltonian
$H=Ap^2+Bx^2$ are
\begin{equation}
E_n=(2n+1)\sqrt{AB}\quad(n=0,1,2,\ldots),
\label{e17}
\end{equation}
the energy levels of $\tilde{H}$ in (\ref{e16}) are
\begin{equation}
E_{m,n}=(2m+1)\sqrt{1+\epsilon}+(2n+1)\sqrt{1-\epsilon}+\frac{1}{1-\epsilon^2}.
\label{e18}
\end{equation}
This result is independent of the constants $a$ and $b$ as well as the choice of
sign of $\eta$. The energy diverges at the critical value $|\epsilon|=
1$ and becomes complex for $|\epsilon|>1$. Thus, there are two regions,
depending on whether the Hermitian component of the Hamiltonian is coupled
strongly or weakly to the non-Hermitian component of the Hamiltonian.

This result is not specific to the choice of coefficients in (\ref{e8}), as we
now show by considering the more general Hamiltonian
\begin{equation}
H=(p^2+\omega_1^2x^2)+(q^2+\omega_2^2y^2+2i\lambda y)+2\epsilon xy.
\label{e19}
\end{equation}
Again, we take the $\cQ$ operator to have the form ${\cQ}=\alpha p+\beta q$, and
determine $\alpha$ and $\beta$ by the condition that $H^\dag=e^{-\cQ}He^{\cQ}$.
We obtain
\begin{equation}
\alpha=\frac{\epsilon\lambda}{\omega_1^2\omega_2^2-\epsilon^2}\quad\hbox{and}
\quad\beta=-\frac{\omega_1^2\lambda}{\omega_1^2\omega_2^2-\epsilon^2}.
\label{e20}
\end{equation}
The equivalent Hermitian Hamiltonian is then
\begin{equation}
\tilde{H}=p^2+\omega_1^2 x^2+q^2+\omega_2^2 y^2+2\epsilon xy+\frac{\lambda^2
\omega_1^2}{\omega_1^2\omega_2^2-\epsilon^2}.
\label{e21}
\end{equation}

Making the {\it ansatz} in (\ref{e13}), we now obtain the unknown coefficients
$a$ through $h$ by following the same procedure as above. The solutions are:
\begin{eqnarray}
c=\gamma a,\quad d=-\frac{b}{\gamma},\quad e=\frac{1}{a(1+\gamma^2)},\nonumber\\
f=\frac{\gamma^2}{b(1+\gamma^2)},\quad g=\frac{\gamma}{a(1+\gamma^2)},\quad
h=-\frac{\gamma}{b(1+\gamma^2)},
\label{e22}
\end{eqnarray}
where $\gamma$ satisfies the quadratic equation
\begin{equation}
\epsilon\gamma^2+2D\gamma-\epsilon=0
\label{e23}
\end{equation}
and we define
\begin{equation}
D=\half(\omega_1^2-\omega_2^2)\quad\hbox{and}\quad S=\half(\omega_1^2+
\omega_2^2).
\label{e24}
\end{equation}

The resulting Hermitian Hamiltonian $\tilde{H}$ is
\begin{equation}
\tilde{H}=P^2+\Omega_1^2 X^2+Q^2+\Omega_2^2 Y^2 +\frac{\lambda^2\omega_1^2}
{\omega_1^2\omega_2^2-\epsilon^2},
\label{e25}
\end{equation}
where we have used the freedom in the choice of $a$ and $b$ to set $a^2=1/(1+
\gamma^2)$ and $b^2=\gamma^2/(1+\gamma^2)$ and where the parameters $\Omega_1$
and $\Omega_2$ are given by
\begin{equation}
\Omega_1^2=S\pm\sqrt{D^2+\epsilon^2}\quad\hbox{and}\quad
\Omega_2^2=S\mp\sqrt{D^2+\epsilon^2}.
\label{e26}
\end{equation}
The energy levels of the Hamiltonian (\ref{e25}) are
\begin{eqnarray}
E_{m,n}=(2m+1)\sqrt{S+\left(D^2+\epsilon^2\right)^{1/2}}\nonumber\\
\qquad\qquad+(2n+1)\sqrt{S-\left(D^2+\epsilon^2\right)^{1/2}}+\frac{\lambda^2
\omega_1^2}{\omega_1^2\omega_2^2-\epsilon^2}.
\label{e27}
\end{eqnarray}
Again, we find that the energy diverges, this time at the critical value
$|\epsilon|=\omega_1\omega_2$, and for $|\epsilon|$ larger than this value the
energy becomes complex. Thus, again there are two regions, depending on whether
the Hermitian component of the Hamiltonian is coupled strongly or weakly to the
non-Hermitian component of the Hamiltonian.

\subsection{Two Coupled Shifted Simple Harmonic Oscillators}

The pattern that we observed in the previous subsection re-emerges when we
consider two coupled $\cP\cT$-symmetric non-Hermitian Hamiltonians:
\begin{equation}
H=(p^2+x^2+2i\lambda x)+(q^2+y^2+2i\mu y)+2\epsilon xy.
\label{e28}
\end{equation}
As before, we choose  ${\cQ}=\alpha p+\beta q$, and determine $\alpha$ and
$\beta$ by condition (\ref{e6}). This gives
\begin{equation}
\alpha=\frac{\epsilon\mu-\lambda}{1-\epsilon^2}\quad\hbox{and}\quad
\beta=\frac{\epsilon\lambda-\mu}{1-\epsilon^2}.
\label{e29}
\end{equation}
Applying (\ref{e7}), we obtain the equivalent Hermitian Hamiltonian
\begin{equation}
\tilde{H}=e^{-\cQ/2}He^{\cQ/2}=p^2+x^2+q^2+y^2+2\epsilon xy
+\frac{\lambda^2+\mu^2-2\epsilon\lambda\mu}{1-\epsilon^2},
\label{e30}
\end{equation}
which is exactly the same as the result in (\ref{e12}), apart from the
additive constant.

\subsection{Simple Harmonic Oscillator Coupled to Swanson Hamiltonian}
Here we consider the non-Hermitian system described by the Swanson Hamiltonian
\cite{Swanson}, written in terms of coordinate and momentum variables instead of
creation and annihilation operators:
\begin{equation}
H=(p^2+x^2)+(q^2+y^2+ic\{q,y\}_+)+2\epsilon xy.
\label{e31}
\end{equation}
We can exploit the ambiguity in $\cQ$ for the Swanson Hamiltonian itself to
choose $\cQ=-cy^2$. This shifts $q\to q-icy$ but leaves $y$, and hence the
coupling term $2\epsilon xy$, unchanged. The equivalent Hermitian Hamiltonian is
then
\begin{equation}
\tilde{H}=p^2+x^2+q^2+(1+c^2)y^2+2\epsilon xy,
\label{e32}
\end{equation}
which can be diagonalized to give
\begin{equation}
h=P^2+\Omega_1^2 X^2+Q^2+\Omega_2^2Y^2,
\label{e33}
\end{equation}
where
\begin{equation}
\Omega_{1,2}^2=1+\half c^2\left(1\pm\sqrt{1+4\epsilon^2/c^4}\right).
\label{e34}
\end{equation}

Notice that the eigenvalues
\begin{equation}
E_{m,n}=(2m+1)\Omega_1+(2n+1)\Omega_2
\label{e35}
\end{equation}
now become complex when $\epsilon^2>1+c^2$. Indeed, in all of the examples
studied in this section, the overall Hamiltonians are $\cP\cT$ symmetric, and
the transition to complex eigenvalues is a signal of the spontaneous breakdown
of that symmetry.

\section{Coupling to Generic Hermitian Hamiltonian}
\label{s4}
In this section we examine the physical system described by a non-Hermitian $\cP
\cT$-symmetric harmonic oscillator Hamiltonian $H_1=p^2+x^2+2ix$ coupled to a
general Hermitian Hamiltonian $H_2=p^2+V(y)$. The only assumptions we will make
are that $H_2$ is separately $\cP$ and $\cT$ symmetric. Thus, we assume that
$V(y)$ is real and is symmetric under parity reflection: $V(y)=V(-y)$. Using
only these assumptions we are able to show that the perturbative expansion for
the ground-state energy is real up to ${\rm O}(\epsilon^2)$.

The ground-state eigenfunction
\begin{equation}
\eta(x)=e^{-(x+i)^2/2}
\label{e36}
\end{equation}
of the $\cP\cT$-symmetric harmonic oscillator satisfies the Schr\"odinger
equation
\begin{equation}
-\eta''(x)+(x^2+2ix)\eta(x)=2\eta(x),
\label{e37}
\end{equation}
whose ground-state energy is $2$. Denoting the ground-state energy of $H_2$ by
$\Lambda$, the ground-state wave function $\psi(y)$ satisfies the Schr\"odinger
equation
\begin{equation}
-\psi''(y)+V(y)\psi(y)=\Lambda\psi(y).
\label{e38}
\end{equation}

We couple $H_1$ and $H_2$ via the coupling term $\epsilon xy$, so that the total
Hamiltonian is $H=H_1+H_2+\epsilon xy$, with $\epsilon$ considered as a small
parameter. The ground-state eigenfunction $\Phi(x,y)$ of the combined system
then satisfies the Schr\"odinger equation
\begin{equation}
-\Phi_{xx}+(x^2+2ix)\Phi-\Phi_{yy}+V(y)\Phi+\epsilon xy\Phi=E\Phi.
\label{e39}
\end{equation}

Let us calculate $E$ and $\Phi(x,y)$ perturbatively. The first three terms in
the perturbation expansion for the energy are
\begin{equation}
E=2+\Lambda+\epsilon E_1 +\epsilon^2 E_2+\ldots
\label{e40}
\end{equation}
and we write
\begin{equation}
\Phi(x,y)=\Phi_0(x,y)+\epsilon\Phi_1(x,y)+\epsilon^2\Phi_2(x,y)+\ldots,
\label{e41}
\end{equation}
where
\begin{equation}
\Phi_0(x,y)=e^{-(x+i)^2/2}\psi(y).
\label{e42}
\end{equation}

The coefficient of $\epsilon^1$ in the expansion of Eq.~(\ref{e39}) is
\begin{equation}
-(\Phi_1)_{xx}+(x^2+2ix)\Phi_1-(\Phi_1)_{yy}+V(y)\Phi_1=-xy\Phi_0+E_0\Phi_1+
E_1\Phi_0.
\label{e43}
\end{equation}
The solution to the homogeneous part of this equation is satisfied by $\Phi_0(x,
y)=\eta(x)\psi(y)$. Using the method of reduction of order, we therefore set
$\Phi_1=\Phi_0(x,y)Q(x,y)$. The integrating factor of the resulting equation is
$\Phi_0$. Multiplying by this integrating factor gives the differential equation
\begin{equation}
(\Phi_0^2Q_x)_x+(\Phi_0^2Q_y)_y=(-xy+E_1)\Phi_0^2.
\label{e44}
\end{equation}

To find $E_1$ we integrate this equation over all $x$ and $y$ and note that the
integrals over the total derivatives vanish. This gives the following expression
for $E_1$:
\begin{equation}
E_1=\frac{\int_{-\infty}^\infty dx\,x e^{-(x+i)^2}\int_{-\infty}^\infty dy\,y
\phi^2(y)}{\int_{-\infty}^\infty dx\,e^{-(x+i)^2}\int_{-\infty}^\infty dy\,
\phi^2(y)}.
\label{e45}
\end{equation}
The integral over $\phi^2(y)$ in the numerator vanishes because of parity
symmetry. Thus $E_1=0$. This result simplifies the differential equation
satisfied by $Q(x,y)$ to:
\begin{equation}
(\Phi_0^2 Q_x)_x+(\Phi_0^2 Q_y)_y=-xy\Phi_0^2.
\label{e46}
\end{equation}

Proceeding to next order, we find the coefficient of $\epsilon^2$ in the
expansion of Eq.~(\ref{e39}):
\begin{equation}
-(\Phi_2)_{xx}+(x^2+2ix)\Phi_2-(\Phi_2)_{yy}+V(y)\Phi_2=-xy\Phi_1+E_0\Phi_2+
E_2\Phi_0.
\label{e47}
\end{equation}
To solve this equation we again use reduction of order and set $\Phi_2(x,y)=
\Phi_0(x,y)R(x,y)$. Multiplying by the integrating factor $\Phi_0$, we get
\begin{equation}
(\Phi_0^2R_x)_x+(\Phi_0^2R_y)_y=(-xyQ+E_2)\Phi_0^2.
\label{e48}
\end{equation}
The next correction to the energy comes from integrating this equation over all
space:
\begin{equation}
E_2=\frac{\int_{-\infty}^\infty dx\int_{-\infty}^\infty dy\,xyQ(x,y)\Phi_0^2
(x,y)}{\int_{-\infty}^\infty dx\int_{-\infty}^\infty dy\,\Phi_0(x,y)^2}.
\label{e49}
\end{equation}
This time the correction does not vanish and we therefore must determine whether
it is real or complex. Note that the integral in the denominator is real.

Let us assume that the numerator $I\equiv \int_{-\infty}^\infty dx\int_{-\infty}
^\infty dy\,xyQ(x,y)\Phi_0^2(x,y)$ is complex and expand $\Phi_0^2(x,y)$ into
its real and imaginary parts:
\begin{equation}
\Phi_0^2(x,y)=\psi^2(y)e^{1-x^2}\left[\cos(2x)-i\sin(2x)\right]\equiv E(x,y)+i F
(x,y).
\label{e50}
\end{equation}
Note that $E$ is even in $x$ and $y$, while $F$ is odd in $x$ and even in $y$.
Next, we let $Q=S+iT$. Then (\ref{e46}) reads
\begin{equation}
\left[(E+iF)(S_x+iT_x)\right]_x +\left[(E+iF)(S_y+iT_y)\right]_y=xy(E+iF).
\label{e51}
\end{equation}
Taking the real and imaginary parts of this equation, we obtain
\begin{equation}
(ES_x-FT_x)_x+(ES_y-FT_y)_y=xyE
\label{e52}
\end{equation}
for the real part, and
\begin{equation}
(ET_x+FS_x)_x+(ET_y+FS_y)_y=xyF
\label{e53}
\end{equation}
for the imaginary part. We conclude that $S$ is odd in $x$ and $y$ and that $T$
is even in $y$ and odd in $x$. Now
\begin{equation}
{\rm Im}(I)=\int_{-\infty}^\infty dx\int_{-\infty}^\infty dy\,xy(ET+SF)\ .
\end{equation}
Here $ET$ is even in $y$, while $FS$ is even in $x$. Thus, we have shown that
$I$ is real.

This result establishes that for small enough $\epsilon$ the ground-state energy
remains real. However, we do not have sufficient information to determine
whether there is a critical value of $\epsilon$ at which the energy becomes
complex. If, as in Ref.~\cite{Herglotz}, we were able to establish that $E$ is
a {\it Herglotz} function of $\epsilon$, that is, that ${\rm Im}(E)$ has the
same sign as ${\rm Im}(\epsilon)$, then we would indeed know that there was such
a critical value because a Herglotz function that is entire must be linear
\cite{BO}, whereas we have shown that $E_2\neq 0$. Unfortunately we are at the
moment unable to construct a proof of the Herglotz property of $E$.

\section{Summary}
\label{s5}

We have shown in a number of examples that it is possible to couple Hermitian
and $\cP\cT$-symmetric non-Hermitian Hamiltonians together in such a way that
the energy eigenvalues of the combined system remain real for sufficiently small
values of the coupling $\epsilon$. In the matrix model and in all of the
quadratic systems we have studied there is a critical range of the coupling,
which, if exceeded, results in a complex spectrum. For coupling to a more
generic $\cP\cT$-symmetric potential we have as yet no analytic proof of the
existence of a critical point in $\epsilon$.

\vspace{0.5cm}
\footnotesize
\noindent
CMB thanks N.~Kaloper for discussions regarding coupled $\cP\cT$-symmetric
quantum systems. As an Ulam Scholar, CMB receives financial support from the
Center for Nonlinear Studies at the Los Alamos National Laboratory. CMB is also
supported by a grant from the U.S. Department of Energy. HFJ is grateful for the
hospitality of the Center for Nonlinear Studies at the Los Alamos National
Laboratory.

\end{document}